\documentclass[twocolumn,showpacs,preprintnumbers,
superscriptaddress]{revtex4}

\renewcommand\a{\alpha}
\renewcommand\b{\beta}

\newcommand\m{\mu}
\newcommand\n{\nu}

\newcommand\ra{\rightarrow}

\newcommand{\ovl}[1]{\overline{#1}}

\newcommand{\non}{\nonumber\\}

\newcommand{\bee}{\begin{equation}}
\newcommand{\eee}{\end{equation}}
\newcommand{\bea}{\begin{eqnarray}}
\newcommand{\eea}{\end{eqnarray}}
\newcommand{\ba}[1]{\begin{array}{#1}}
\newcommand{\ea}{\end{array}}

\newcommand{\eqrf}[1]{Eq.\ (\ref{#1})}

\begin{document}


\title{Kinetic Equation for Gluons in
the Background Gauge of QCD}

\author{Q. Wang}
\email{qwang@th.physik.uni-frankfurt.de}
\affiliation{
Institute f\"ur Theoretische Physik, J. W.
Goethe-Universit\"at, D-60054 Frankfurt, Germany
}
\affiliation{
Physics Department, Shandong University, 
Jinan, Shandong 250100, People's Repbulic of China
}
\author{K. Redlich}
\email{Krzysztof.Redlich@cern.ch}
\affiliation{
Theory Division, CERN, CH-1211 Geneva 23, Switzerland
}
\affiliation{
Institute for Theoretical Physics, University of Wroclaw,
PL-50204  Wroclaw, Poland
}
\author{H. St\"ocker}
\email{stoecker@th.physik.uni-frankfurt.de}
\author{W. Greiner}
\email{greiner@th.physik.uni-frankfurt.de}
\affiliation{
Institute f\"ur Theoretische Physik,
J. W. Goethe-Universit\"at, D-60054 Frankfurt, Germany
}

\date{\today}

\begin{abstract}

We derive the quantum kinetic equation for a pure gluon plasma, 
applying the background field and closed-time-path method. 
The derivation is more general and transparent than 
earlier works. A term in the equation is found which, 
as in the classical case, 
corresponds to the color charge precession 
for partons moving in the gauge field. 

\end{abstract}


\maketitle

Heavy ion collisions at ultra-relativistic energies are widely
expected to be a laboratory to study the formation and properties
of highly excited QCD matter in the deconfined Quark-Gluon
Plasma (QGP) phase \cite{rev1}. The QGP is considered as a
partonic system being at (or close to) local thermal equilibrium.
Thus, to study the conditions for the possible formation of QGP in
heavy ion collisions one needs to address the question of
thermalization of the initially produced partonic medium
\cite{baier}. In the theoretical aspect this requires the
formulation of the kinetic equations \cite{jp} involving color degrees
of freedom and the non-Abelian structure of QCD dynamics.
Various models for the initial conditions in ultra-relativistic
heavy ion collisions suggest that at the early stage the medium is
dominated by gluon degrees of freedom \cite{sat}. The kinetic 
equation for a pure gluon plasma is thus of special interest.

The usual treatment of the gluon transport equation is based on
the decomposition of the gluon field into a mean field and a
quantum fluctuation. Under this
approximation the gluon transport equation then describes the
kinetics of the quanta in the classical mean field
\cite{heinz,elze88,geiger}. This picture is somewhat similar to the one
used in studying the energy loss of a fast parton 
moving in the soft mean field \cite{mgxw,baier1}. 
To include the classical chromofield into QCD in a proper way, 
one uses the background field method of QCD (BG-QCD) introduced by
DeWitt and 't Hooft \cite{dewitt,thooft,abb81}. 
The advantage of BG-QCD is that it is formulated 
in an explicit gauge invariant manner.

One of the first attempts to derive the gluon transport equation
in BG-QCD was presented in Ref. \cite{elze90}. But the obtained
equation is not transparent. The most recent work has
been done by Blaizot and Iancu \cite{blaizot,blaizot1} 
in the close-time-path (CTP) formalism. There, however,
the authors focus on formulating the transport equation in the
vicinity of equilibrium.

In this paper, we use the CTP and BG-QCD method to derive the kinetic 
equation for the gluon plasma. Our derivation is going
beyond previous results as it is quite general and transparent. 
One of the most important features of 
our work is that a term in the obtained equation 
is shown to correspond to the color charge precession term in 
the classical equation \cite{heinz,wong,elze86,jalilian,kelly}.  

In the following we use $g_{\m\n}={\rm diag}(1,-1,-1,-1)$ as the
metric tensor, and for elegance of the formula 
we always write Lorentz indices as subscripts and
color ones as superscripts for the relevant quantities. 
For the gauge field and its strength tensor we denote 
$A_{\m}\equiv A_{\m}^aT^a$ and 
$F_{\m\n}[A]\equiv F_{\m\n}^a[A]T^a$, 
where $(T^a)^{ij}=if^{iaj}$ are the generators
of the $SU(3)_c$ adjoint representation.
The two-point Green function (GF) or self-energy (SE) are treated
as matrices, so their color and/or Lorentz indices are
sometimes omitted.

Applying the background field method, we decompose the
conventional gluon field into the sum of 
a classical background part $A$ and a quantum fluctuation $Q$. 
Including the appropriate gauge fixing and 
ghost terms for the background gauge 
$D_{\mu}^{ij}[A]Q_{\mu}^j=0$, 
the BG-QCD Lagrangian reads \cite{abb81}
\bea {\cal L}&=&
-\frac 14F_{\mu \nu}^i[A+Q]F_{\mu \nu}^i[A+Q]
-\frac {1}{2\alpha} (D_{\mu}^{ij}[A]Q_{\mu}^j)^2 \non
&&+\overline{C}^i D_{\mu}^{ij}[A]D_{\mu}^{jk}[A+Q]C^k \;,
\label{s} \eea
where $D_{\mu}^{ij}[A(x)]= [\partial _{x\mu}-igA_{\mu}(x)]^{ij}$
is the covariant derivative, $C^i$/$\overline{C}^i$ are the
ghost/anti-ghost field and $\alpha$ is 
the gauge fixing parameter.

The above Lagrangian is invariant under the local gauge
transformation of type I 
(type II transformation is irrelevant to the 
current problem) where the background field transforms as
a conventional gauge fields, 
$A'_{\mu}=UA_{\mu}U^{-1} +ig^{-1} U\partial _{\mu}U^{-1}$, 
while the gluon and ghost fields transform
like a matter field, $Q'_{\mu}=UQ_{\mu}U^{-1}$ \cite{abb81}. 
Here $U(x)=\exp [ig\omega ^a(x)T^a]$ is the transformation matrix.

The non-equilibrium dynamics is usually described in the CTP
formalism \cite{ctp}. Here the action can be written as
$S_{\rm CTP}=S_+-S_-+K(A_{\pm},Q_{\pm})$ 
where all fields in $S_{\pm}$
are defined on the positive/negative time branches and
$K(A_{\pm},Q_{\pm})$ is the kernel incorporating initial state
correlations. The role of the non-local kernel $K(A_{\pm},Q_{\pm})$ 
in $S_{\rm CTP}$ is discussed in Ref.\ \cite{qwang} 
and its coonection with the pinch singularity \cite{pinch} 
will be discussed elsewhere. 
The GF for the gluon has four components: 
$ [ G^{++}, G^{+-}, G^{-+}, G^{--} ]
\equiv [ G^{F}, G^{<}, G^{>}, G^{\ovl{F}} ]$, 
marked by the positive or the negative time branch,
respectively. One can also express the GF 
in the physical representation, where 
the GF is expressed by the symmetric (C), 
retarded (R) and advanced (A) components. 
The two representations are related by a unitary transformation. 
The same expressions for the SE are also valid. 


In the CTP formalism, choosing the physical representation 
for the GF and SE,  
the Dyson-Schwinger equation (DSE) for $G^C$ reads
\bea {\cal D}(x_1)G^C(x_1,x_2) &=&-\int d^4x' 
\big[ \Pi ^C(x_1,x')G^A(x',x_2) \non 
&& +\Pi ^{R}(x_1,x')G^{C}(x',x_2)\big] \;,
\label{sch-dy-c1}\\
G^C(x_1,x_2){\cal D}^{\dagger}(x_2) 
&=&-\int d^4x' 
\big[ G^R(x_1,x')\Pi ^C(x',x_2) \non 
&&+G^{C}(x_1,x')\Pi ^{A}(x',x_2)\big] \;,
\label{sch-dy-c2} \eea
where the differential operators ${\cal D}$ and 
${\cal D}^{\dagger}$ in the Feynman gauge 
\footnote{Due to gauge
conditions: $D^{ij}_{\mu}[A(x_1)]G_{\mu\nu}^{jk}(x_1,x_2)=0$ and
$G^{ij}_{\mu\nu}(x_1,x_2)D_{\nu}^{\dagger ;jk}[A(x_2)]=0$, 
Eqs.\ (\ref{sch-dy-c1},\ref{sch-dy-c2}) are independent of the gauge
parameter $\a $.} 
($\alpha =1)$ are expressed by
\bea
{\mathcal D}^{hi}_{\rho\sigma}
&=&g_{\rho\sigma}D_{\mu}^{ha}[A]D_{\mu}^{ai}[A]
+2gf^{hai}F^a_{\rho\sigma}[A]\;,
\label{d1}\\
{\mathcal D}^{\dagger;hi}_{\rho\sigma}
&=&g_{\rho\sigma}D_{\mu}^{\dagger;ha}[A]D_{\mu}^{\dagger;ai}[A]
+2gf^{hai}F^a_{\rho\sigma}[A] \;,
\label{d2} 
\eea
where $D_{\mu}^{\dagger;ij}[A(x)] \equiv
[ \stackrel{\leftarrow}{\partial }_{x\mu}
+igA_{\mu}(x) ]^{ij}$
is the conjugate covariant derivative
where the differential operator acts
on the function in its left.

In the evolution of the gluonic system one distinguishes different
scales, which characterize quantum and soft collective motion. 
We introduce a mass parameter, $\m$, as the separation point of the
quantum and the kinetic scale. In the weak coupling limit 
$g\ll 1$, the scale of collectivity $\sim 1/(g\mu)$ is much larger than
the typical extension of hard quantum fluctuations $\sim 1/\mu$.
The effect of the classical field $A$ on the hard quanta involves
the coupling $gA$ to the hard propagator and is of the size of the
soft wavelength $\sim 1/(g\mu)$. The above separation of scales is
the basis for the gradient expansion where one expresses all
2-point GFs in terms of the relative $y$ and the central $X$
coordinate. Here are some typical scales: $y=x_1-x_2\sim 1/\mu$,
$X=(x_1+x_2)/2\sim 1/(g\mu)$, $A(X)\sim \mu$ and 
$F[A(X)]\sim g\mu^2$.

In order to obtain the gauge-covariant
kinetic equation one uses the gauge-covariant Wigner function
$\tilde{G}(X,y)$ ($\tilde{G}^C$, $\tilde{G}^>$ or $\tilde{G}^<$)
defined by
\bea
\label{tg-g}
G(x_1,x_2)&=&V(x_1,X)\tilde{G}(X,y)V(X,x_2)\;,
\eea
where $V(z_1,z_2)=T_{\rm P}
\exp (ig\int _{{\rm P};z_2}^{z_1} dz_{\mu}A_{\mu})$
denotes a Wilson link with respect to
the classical background field.
One can also define the Wilson
link as a functional of $A+Q$,
but it is a much more complicated case and
beyond the scope of this work.

The covariant Wigner function $\tilde{G}(X,y)$ transforms as
$U(X)\tilde{G}(X,y)U^{-1}(X)$ where only the transformation at a
single point $X$ is relevant. For $G(x_1,x_2)$, however, the
gauge transformation involves two points and therefore is not
gauge-covariant.

The DSE given by Eqs.\ (\ref{sch-dy-c1}) and (\ref{sch-dy-c2}) can
be expressed in terms of the covariant Wigner functions. 
To evaluate ${\mathcal D}(x_1)G(x_1,x_2)$ 
one needs to know the variation of a Wilson link 
caused by that of its ending points. 
Following Eq.\ (3.15) for $\delta V(z_1,z_2)$ 
in Ref. \cite{elze86}
[note the opposite sign convention for $g$], 
one finds
\bea
D_{x_1\nu}(V_1\tilde{G}V_2)&\approx &V_1
\Big[ (\partial _{y\nu}\tilde{G}) 
+\frac 12 (D_{X\nu}\tilde{G}) 
+ig \frac 12 \tilde{G} A_{\nu}
\non &&
-ig \frac 38 y_{\lambda} F_{\lambda\nu}\tilde{G}
-ig\frac 18 \tilde{G} y_{\lambda} F_{\lambda\nu} \Big] V_2\;,
\label{dx1nu}\\
D^2_{x_1}(V_1\tilde{G}V_2)
& \approx & V_1\Big[ 
\frac 14 (D^2_{X}\tilde{G})
+(\partial _{y}\cdot D_X\tilde{G})
+(\partial ^2_{y}\tilde{G})
\non &&
-ig\frac 34 y_{\lambda} F_{\lambda\nu}
(\partial _{y\nu}\tilde{G})
-ig\frac 14 (\partial _{y\nu}\tilde{G}) y_{\lambda}F_{\lambda\nu}
\non &&
+ig(\partial _{y\nu}\tilde{G})A_{\nu}
-ig\frac 38 y_{\lambda}F_{\lambda\nu}(D_{X\nu}\tilde{G})
\non &&
+ig\frac 12 (D_{X\nu}\tilde{G})A_{\nu} 
+ig\frac 14 \tilde{G}(\partial _{X\nu}A_{\nu})
\non &&
-ig\frac 18 (D_{X\nu}\tilde{G})y_{\lambda}F_{\lambda\nu} 
-\frac 14 g^2 \tilde{G}A^2 \Big] V_2\;,
\nonumber \\
\label{dx2nu}
\eea
where we have used: 
$V_{1}\equiv V(x_1,X)$, $V_{2}\equiv V(X,x_2)$, 
$\tilde{G}\equiv \tilde{G}(X,y)$, 
$D_{X\nu}\equiv D_{\nu}[A(X)]$, 
$D_{t}^2\equiv D_{t\nu}D_{t\nu}$, 
$A_{\m}\equiv A_{\m}(X)$, 
$A^2\equiv A_{\m}(X)A_{\m}(X)$ and 
$F_{\m\n}\equiv F_{\m\n}(X)$. The conjugate expressions
$(V_1\tilde{G}V_2)D^{\dagger}_{x_2\n}$ and
$(V_1\tilde{G}V_2)D^{\dagger 2}_{x_2}$ can be obtained from
Eqs.\ (\ref{dx1nu}) and (\ref{dx2nu}) by taking their Hermitian
conjugates and then interchanging $x_1$ and $x_2$. Using these
results together with \eqrf{tg-g} we derive the following gauge
conditions for $\tilde{G}$ 
\footnote{To derive these results one
needs the following gauge conditions for $G$ ($G^>$, $G^<$ or
$G^C$) with respect to $x_1$ and $x_2$:
$D_{\mu}[A(x_1)]G_{\mu\nu}(x_1,x_2)=0$ and
$G_{\mu\nu}(x_1,x_2)D_{\nu}^{\dagger}[A(x_2)]=0$. They are
consequences of the background gauge condition
$D_{\mu}[A(x)]Q_{\mu}(x)=0$.}:
\begin{equation}
\partial _{X\mu}\tilde{G}_{\mu\nu}
+ig[\tilde{G}_{\mu\nu},A_{\mu}]=0\;,\;\;
\partial _{y\mu}\tilde{G}_{\mu\nu}=0\;,
\label{ggcond}
\end{equation}
which are valid up to $O(g\m)$. The second equation is a
transversality condition for $\tilde{G}_{\mu\nu}$.

Taking the difference between \eqrf{sch-dy-c1} and
\eqrf{sch-dy-c2} and using \eqrf{dx2nu} and its conjugate
expression, we derive the following kinetic equation for the gluon
plasma:
\bea
&& q\cdot \partial _X\tilde{G}_{\alpha\gamma}^{C}
+ig(\tilde{G}_{\alpha\gamma}^{C}q\cdot A -q\cdot
A\tilde{G}_{\alpha\gamma}^{C}) \non
&&+\frac 12 gq_{\nu}F_{\nu\lambda}(\partial _{q\lambda}
\tilde{G}_{\alpha\gamma}^{C})
+\frac 12 g(\partial _{q\lambda}\tilde{G}_{\alpha\gamma}^{C})
q_{\nu}F_{\nu\lambda}\non
&&+g(F_{\alpha\beta}\tilde{G}_{\beta\gamma}^{C}
-\tilde{G}_{\alpha\beta}^{C}F_{\beta\gamma})=0\;,
\label{treq1}
\eea
where  $\tilde{G}_{\alpha\gamma}^{C}\equiv
\tilde{G}_{\alpha\gamma}^{C}(X,q)$.

The above equation is located at the collective coordinate $X$ and
is {\it gauge-covariant} under the local gauge transformation
$U(X)$, i.e. it transforms as $U(\cdots )U^{-1}$. Indeed, noting
that both $F_{\m\n}$ and $\tilde{G}_{\alpha\gamma}^{C}$ are 
gauge-covariant and $\partial_{q\m}$ does not affect $U$, 
it is obvious that the last three terms are gauge-covariant. 
To verify that the
first two terms also preserve gauge covariance, we explicitly
write down their transformation 
\bea q\cdot \partial
_X\tilde{G}_{\alpha\gamma}^{C} &\ra & U(q\cdot \partial_X
\tilde{G}_{\alpha\gamma}^{C})U^{-1} +(q\cdot \partial_X
U)\tilde{G}_{\alpha\gamma}^{C}U^{-1} \non
&&+U\tilde{G}_{\alpha\gamma}^{C}q\cdot \partial_X U^{-1}\;, \non
ig\tilde{G}_{\alpha\gamma}^{C}q\cdot A &\ra &
igU\tilde{G}_{\alpha\gamma}^{C}q\cdot A\; U^{-1}
-U\tilde{G}_{\alpha\gamma}^{C}q\cdot \partial_X U^{-1}\;, \non -ig
q\cdot A \tilde{G}_{\alpha\gamma}^{C} &\ra & -ig U q\cdot A
\tilde{G}_{\alpha\gamma}^{C} U^{-1} \non && - ( q\cdot \partial_X
U ) \tilde{G}_{\alpha\gamma}^{C} U^{-1}. 
\nonumber 
\eea 
With the above gauge transformations 
one can clearly see that the sum of the
first two terms in \eqrf{treq1} indeed transforms as 
$U(\cdots )U^{-1}$ and therefore preserves the gauge covariance.

The quantum kinetic equation 
(\ref{treq1}) is derived in a
quite general and consistent manner 
in BG-QCD and CTP formalism. 
No further approximations or requirements going
beyond the gradient expansion 
were used. We notice that a result 
similar to Eq.\ (\ref{treq1}) was 
also obtained in Ref.\ \cite{mrow}.
There, however, the kinetic equation was derived
by making the gradient expansion of the equation of motion 
for the Wigner function (not in CTP formalism). 
In addition the derivation was made 
in the fundamental (not the adjoint) color 
space in QCD (not BG-QCD). Finally the assumption
was made in Ref.\ \cite{mrow} 
that the Wigner function is proportional to the quadratic
product of the generators of the fundamental 
representation. The approach presented here 
is quite general and does not require
any specific assumptions on the structure
of the Wigner function.



In the following we will compare \eqrf{treq1} with 
the classical equation. Especially we will show the 
physical connection of the second term of \eqrf{treq1} 
to the color precession. The classical kinetic equation 
for the color singlet distribution
function $f(x,p,Q)$ has the following form 
\cite{heinz,wong,elze86,jalilian,kelly}:
\bea
\label{wongeq} 
&&p_{\mu}[\partial _{\mu}-gQ^aF_{\mu\nu}^a
\partial _{p\nu} \non
&&-gf^{abc}A_{\mu}^b(x)Q^c\partial _{Q^a}] f(x,p,Q)=0\;,
\eea
where $Q^a$ is the classical color charge and  $a=1,.., N^2_c-1$.
The last term of \eqrf{wongeq} describes the rotation or 
precession of the color charge \cite{heinz,wong,elze86}. 


Comparing \eqrf{wongeq} with the quantum expression \eqrf{treq1}
it is clear that the color singlet distribution function $f$ is
replaced by the gauge-covariant Wigner function $\tilde{G}^C$
which is a color matrix in the adjoint representation. One can
also recognize that the first, third and fourth terms of
\eqrf{treq1} are the quantum generalization of the first two terms
in \eqrf{wongeq}. The last term in \eqrf{treq1} appears from the
covariant operators \footnote{This term can be written in
different form by using generators of the Lorentz transformation
in {\em vector} representation. Similar term can be found for the
quark, but expressed through generators in {\em spinor}
representation.} and hence is not present in the classical
equation.

Particularly interesting is the appearance of the second term in
\eqrf{treq1}. We have seen that its presence is crucial to assure
the gauge covariance of the Vlasov equation. 
This term has an interesting physical meaning. It is the quantum
analogue to the color charge precession in the classical kinetic
equation. To see this more clearly, one can expand
$\tilde{G}^C_{\a\b}(X,q)$ with respect to the expansion 
parameter $gT^a A^a_{\m}(X)$ from the Wilson link in \eqrf{tg-g}. 
This expansion can be also understood as the result of 
the $AQQ$, $AQQQ$ and $AAQQ$ vertices. Then we have 
\bea
\tilde{G}^C_{\a\b}(X,q)&=&N_{0\a\b}(X,q)+gT^a A^a_{\m}(X) 
N_{1\a\b;\m}(X,q) \non
&&+g^2 T^a T^b A^a_{\m}(X) A^b_{\n}(X) N_{2\a\b;\m\n}(X,q)\non 
&&+\cdots \;, 
\label{expansion} 
\eea
where $T^a$ are quantum analogues to the classical color charges
$Q^a$; $N_{i\a\b}(X,q)$ with $i=0,1,2,\cdots $ 
are color singlet functions. Each term of 
the expansion corresponds to an order of the color 
inhomogeneity in the gluonic medium due to its interaction 
with the background field. If the background field is 
refered to the soft mean field, its magnitude 
should vanish when the system approaches equilibrium. 
Then only the first singlet term survives in \eqrf{expansion}, 
which means the color homogeneity of the gluonic medium. 
This is somewhat similar to the 
multipole expansion for an electromagnetic source 
where the moments of dipole, quadrupole etc. 
describe increasing orders of spatial 
inhomogeneity for the electromagnetic charges. 
In weak coupling, as the lowest order approximation, 
we keep only the first two terms in \eqrf{expansion}. 
Then the second term of \eqrf{treq1} becomes
\begin{equation}
ig(\tilde{G}_{\alpha\gamma}^{C}q\cdot A -q\cdot
A\tilde{G}_{\alpha\gamma}^{C})\simeq -g f^{abc} q_{\m} A^b_{\m}
T^c \partial_{T^a} \tilde{G}_{\alpha\gamma}^{C}
\end{equation}
which reproduces the classical
color precession term in Eq.\ (\ref{wongeq}).

Since we know that $D_X\sim gA(X)\sim g\m$, the first two terms of
\eqrf{treq1}, i.e. $q\cdot \partial _X\tilde{G}_{\alpha\gamma}^{C}$ 
and the color precession term, are
at leading order $O(g\m ^2)$, while other terms are at subleading
order $O(g^2\m ^2)$. In the vicinity of equilibrium the natural
scale in the system is the temperature $T$. The mean distance
between particles is of the order of $\sim 1/T$, while $1/(gT)$
characterizes the scale of collective excitations
\cite{blaizot,blaizot1}. For small coupling constant $g$ these two
scales are well separated. The covariant Wigner functions can be
expanded around their equilibrium values:
$\tilde{G}=\tilde{G}^{(0)}+\delta\tilde{G}$, where the equilibrium
function $\tilde{G}^{(0)}$ is a color singlet and the fluctuation
$\delta\tilde{G}\sim g^2\tilde{G}$. Typical scales are $q\sim T$,
$D_X\sim g^2T$, $gF\sim (D_X)^2\sim g^4T^2$. Thus, at leading
order, only the first term of \eqrf{treq1} survives and the
precession term vanishes due to 
the color-singlet nature of $\tilde{G}^{(0)}$. 
The linearized version of Eq.\ (\ref{treq1}) with 
respect to $\delta \tilde G$
corresponds to the equation formulated in 
the background Coulomb gauge in Ref.~\cite{blaizot}.

The quantum fluctuations near equilibrium were also
considered in Ref.\ \cite{litim} in the context of 
the classical collisionless transport equation. 
It is quite natural to carry out the same study 
from our quantum approach. First, BG-QCD deals 
with the classical field and the quantum fluctuation 
in a systematic way. The quantum field plays the 
similar role to the field fluctuation in Ref.\ \cite{litim}. 
Second, in the quantum approach, 
corresponding to the phase-space distribution, 
we deal with the GFs which can 
be expanded around their equilibrium values 
following Eq.\ (\ref{expansion}). 
One also needs to complete 
the equations by including the field
equation $D_{\m}F_{\m\n}=\langle J_{\n}\rangle$. 
Here the averaged induced current 
$\langle J_{\n}\rangle$ is 
related to the 2- and 3-point GFs \cite{blaizot1}
which finally depend on $\tilde{G}$. 
Thus it can be expanded according to 
Eq.\ (\ref{expansion}) as well.

The analogy and differences of the quantum and the classical
Vlasov equation can also be shown 
by comparing the equations for the color moments.
Corresponding to \eqrf{treq1} one gets
\bea
&& q\cdot \partial _X h_{\alpha\gamma}
+g q_{\nu} F_{\nu\lambda}^a \partial
_{q\lambda} h_{\alpha\gamma}^a
\non
&& +g ( F_{\alpha\beta}^ah_{\beta\gamma}^a
-h_{\alpha\beta}^aF_{\beta\gamma}^a )=0\;,
\label{treq01} \\
&& q\cdot \partial _X h_{\alpha\gamma}^a
+gf^{abc}q\cdot A^bh_{\alpha\gamma}^c
\non
&& +g q_{\nu} F_{\nu\lambda}^b\partial _{q\lambda}
\frac 12[h_{\alpha\gamma}^{ab}+h_{\alpha\gamma}^{ba}]
\non
&& +g(F_{\alpha\beta}^bh_{\beta\gamma}^{ab}
-h_{\alpha\beta}^{ba}F_{\beta\gamma}^b)=0\;,
\label{treq2}
\eea
where we defined $h_{\alpha\gamma}={\rm
Tr}(\tilde{G}_{\alpha\gamma}^{C})$, $h_{\alpha\gamma}^a={\rm
Tr}(T^a\tilde{G}_{\alpha\gamma}^{C})$ and
$h_{\alpha\gamma}^{ab}={\rm
Tr}(T^aT^b\tilde{G}_{\alpha\gamma}^{C})$.
The classical equations for color moments of $f(x,p,Q)$ can be found 
in Ref.\ \cite{heinz}.   
Comparing  Eq.\ (\ref{treq01}) and (\ref{treq2}) 
with the classical expressions in Ref.\ \cite{heinz}, 
one sees that apart from the term 
$(F_{\alpha\beta}h_{\beta\gamma}-h_{\alpha\beta}F_{\beta\gamma})$, 
which comes from the covariant operators, 
the quantum and classical equation have similar
structure. The identification of the color precession term in
Eq.\ (\ref{treq1}) is then straightforward.


In summary, by applying the CTP and BG-QCD formalism we have
derived the kinetic equation for the gluon. 
The derivation is more general and transparent than 
earlier works. The kinetic equation is with respect to
gauge-covariant Wigner function, 
which is a matrix in adjoint color space. 
A notable feature of our work is that a term is found 
which, as in the classical case, 
corresponds to the color precession of the parton. 
This is the non-Abelian analogue 
to the Larmor precession for particles with 
the magnetic moment moving in a magnetic field. 
We see that this term is necessary to the gauge covariance of the
kinetic equation.

One of us Q.W. acknowledges financial support of the Alexander von
Humboldt-Foundation (AvH) and appreciated help from D. Rischke.
This work is partially funded by DFG, BMBF and GSI. K.R.
acknowledges a partial support of the Polish Committee for
Scientific Research (KBN-2P03B 03018). Stimulating comments and
discussions with R. Baier, J.-P. Blaizot, E. Iancu and S. Leupold
are acknowledged. Our special thanks go to X.-N. Wang for his help
and interest in this work.

\end{document}